\documentstyle[12pt]{article}
\newcommand {\beq}{\begin{equation}}
\newcommand {\eeq}{\end{equation}}
\newcommand {\beqa}{\begin{eqnarray}}
\newcommand {\eeqa}{\end{eqnarray}}
\newcommand {\n}{\nonumber \\}

\begin{document}
\setlength{\oddsidemargin}{0cm}
\setlength{\baselineskip}{7mm}

\begin{titlepage}
 \renewcommand{\thefootnote}{\fnsymbol{footnote}}
$\mbox{ }$
\begin{flushright}
\begin{tabular}{l}
KEK-TH-649 \\
KUNS-1608\\
\end{tabular}
\end{flushright}

~~\\
~~\\
~~\\

\vspace*{0cm}
    \begin{Large}
       \vspace{2cm}
       \begin{center}
         {Wilson Loops in Noncommutative Yang Mills}      \\
       \end{center}
    \end{Large}

  \vspace{1cm}

\begin{center}
           Nobuyuki I{\sc shibashi}$^{1)}$\footnote
           {
e-mail address : ishibash@post.kek.jp},
           Satoshi I{\sc so}$^{1)}$\footnote
           {
e-mail address : satoshi.iso@kek.jp},\\
           Hikaru K{\sc awai}$^{2)}$\footnote
           {
e-mail address : hkawai@gauge.scphys.kyoto-u.ac.jp}{\sc and}
           Yoshihisa K{\sc itazawa}$^{1)}$\footnote
           {
e-mail address : kitazawa@post.kek.jp}\\
        $^{1)}$ {\it High Energy Accelerator Research Organization (KEK),}\\
               {\it Tsukuba, Ibaraki 305-0801, Japan} \\
        $^{2)}$ {\it Department of Physics, Kyoto University,
Kyoto 606-8502, Japan}\\
\end{center}

\vfill

\begin{abstract}
\noindent
\end{abstract}
We study the correlation functions of the Wilson loops
in noncommutative Yang-Mills theory
based upon its equivalence to twisted reduced models.
We point out that there is a crossover at the noncommutativity scale.
At large momentum scale, the Wilson loops in
noncommmutative Yang-Mills represent extended objects.
They coincide with those in ordinary Yang-Mills theory in low energy limit.
The correlation functions on D-branes
in IIB matrix model exhibit the identical crossover behavior.
It is observed to be consistent with the supergravity description
with running string coupling.
We also explain that the results of Seiberg and Witten can be
simply understood in our formalism.

\vfill
\end{titlepage}
\vfil\eject

\section{Introduction}
\setcounter{equation}{0}
A large $N$ reduced model has been proposed as a nonperturbative
formulation of type IIB superstring theory\cite{IKKT}\cite{FKKT}.
It is defined by the following action:
\beq
S  =  -{1\over g^2}Tr({1\over 4}[A_{\mu},A_{\nu}][A^{\mu},A^{\nu}]
+{1\over 2}\bar{\psi}\Gamma ^{\mu}[A_{\mu},\psi ]) .
\label{action}
\eeq
Here $\psi$ is a ten dimensional Majorana-Weyl spinor field, and
$A_{\mu}$ and $\psi$ are $N \times N$ Hermitian matrices.
It is formulated in a manifestly covariant way which enables us
to study the nonperturbative issues of superstring theory.
In fact we can in principle predict the dimensionality of spacetime,
the gauge group and the matter contents by solving this model.
We have already initiated such investigations
in \cite{AIKKT}\cite{ISK}.
We refer our recent review for more
detailed expositions and references\cite{review}.
We also note a deep connection between our approach and
noncommutative geometry\cite{AIIKKT}\cite{Connes}\cite{CDS}.

The IIB matrix model is invariant under the
$\cal{N}$=2 supersymmetry:
\beqa
\delta^{(1)}\psi &=& \frac{i}{2}
                     [A_{\mu},A_{\nu}]\Gamma^{\mu\nu}\epsilon ,\n
\delta^{(1)} A_{\mu} &=& i\bar{\epsilon }\Gamma^{\mu}\psi ,
\label{Ssym1}
\eeqa
and
\beqa
\delta^{(2)}\psi &=& \xi ,\n
\delta^{(2)} A_{\mu} &=& 0.
\label{Ssym2}
\eeqa
If we take a linear combination of $\delta^{(1)}$ and $\delta^{(2)}$ as
\beqa
\tilde{\delta}^{(1)}&=&\delta^{(1)}+\delta^{(2)}, \n
\tilde{\delta}^{(2)}&=&i(\delta^{(1)}-\delta^{(2)}),
\eeqa
we obtain the $\cal{N}$=2 supersymmetry algebra.
\beqa
(\tilde{\delta}^{(i)}_{\epsilon}\tilde{\delta}^{(j)}_{\xi}
    -\tilde{\delta}^{(j)}_{\xi}\tilde{\delta}^{(i)}_{\epsilon})\psi   &=&0
,\n
(\tilde{\delta}^{(i)}_{\epsilon}\tilde{\delta}^{(j)}_{\xi}
    -\tilde{\delta}^{(j)}_{\xi}\tilde{\delta}^{(i)}_{\epsilon})A_{\mu}&=&
                                 2i\bar{\epsilon}\Gamma^{\mu}\xi
\delta_{ij}.
\eeqa
The $\cal{N}$=2 supersymmetry is a crucial element of superstring theory.
It imposes strong constraints on the spectra of particles.
Furthermore it determines the structure of the interactions uniquely in
the light-cone string field theory\cite{FKKT}.
The IIB matrix model is a nonperturbative formulation which possesses
such a symmetry. Therefore it has a very good chance to capture the
universality class of IIB superstring theory.
These symmetry considerations force us to interpret
the eigenvalues of $A_{\mu}$ as the space-time coordinates.
Note that our argument is independent of the D-brane interpretations
which are inevitably of semiclassical nature\cite{Polchinski}.

The equation of motion of (\ref{action}) with $\psi=0$ is
\beq
[A_{\mu},[A_{\mu},A_{\nu}]]=0.
\label{SEOM}
\eeq
The cases in which
$[A_\mu,A_\nu] =c-number\equiv C_{\mu\nu}$
have a special meaning.
These correspond to BPS-saturated backgrounds \cite{WO}. Indeed, by setting
$\xi$ equal to $\pm \frac{1}{2}C_{\mu\nu} \Gamma^{\mu\nu}\epsilon$
in the $\cal{N}$=2 supersymmetry (\ref{Ssym1}) and (\ref{Ssym2}),
we obtain the relations
\beqa
(\delta^{(1)}\mp \delta^{(2)}) \psi &=&0 ,\n
(\delta^{(1)}\mp \delta^{(2)}) A_{\mu}&=& 0.
\eeqa
Namely, half of the supersymmetry is preserved in these backgrounds.
The D-branes in IIB matrix model have been investigated in
\cite{olesen}\cite{tseytlin}\cite{ishibashi}\cite{takata}.

The bosonic part of the action vanishes for the commuting matrices
$(A_{\mu})_{ij}=x^{\mu}_i \delta _{ij}$ where $i$ and $j$ are color indices.
These are the generic classical vacuum configurations of the model.
We have proposed to interpret $x^{\mu}_i$ as the space-time coordinates.
If such an interpretation is correct, the distributions of the eigenvalues
determine the extent and the dimensionality
of spacetime. Hence the structure of spacetime is dynamically determined by
the
theory.
As we have shown in \cite{AIKKT}, spacetime exists as a single bunch
and no single eigenvalue can escape from the rest.
However the appearance of a smooth manifold itself is not apparent
in this approach since we find four dimensional fractals in a simple
approximation.
Although it is very plausible that
gauge theory and gravitation may appear as low energy
effective theory, we are still not sure how matter fields propagate
\cite{ISK}.

The situation drastically simplifies if we consider noncommutative
backgrounds. These are the D-brane like solutions which preserve
a part of SUSY.
We can indeed show that gauge theory appears as the low energy effective
theory.
In the case of $m$ coincident D-branes, we obtain noncommutative super-Yang
Mills
theory of 16 supercharges in the gauge group of $U(m)$\cite{AIIKKT}.
It is of course well-known that the low energy effective action
for D-branes is super Yang-Mills theory.
If we mod out the theory with the translation operator,
we immediately find the corresponding super Yang-Mills theory
\cite{BFSS}\cite{WT}.
Noncommutative Yang-Mills theories have been obtained by the
compactification on noncummutative tori\cite{CDS}.
By `compactification', however,
we may modify the theory by throwing away many degrees of freedom.

We have pointed out that well-known twisted reduced models\cite{twisted}
are equivalent to noncommutative Yang-Mills theory.
The expansion around the infinitely extended D-branes in IIB matrix model
defines a twisted reduced model.
Using the equivalence,
we have found noncommutative Yang-Mills theory in IIB matrix model.
Our proposal is that IIB matrix model with D-brane backgrounds
provides us a concrete definition of noncommutative Yang-Mills theory.

In this paper we further investigate twisted reduced models
as noncommutative Yang-Mills theory. We investigate the
correlation functions of gauge invariant operators namely Wilson loops.
It has been well known that the twisted reduced model is equivalent
to large $N$ gauge theory. The natural question to arise is
how to reconcile the noncommutative Yang-Mills interpretation
with the large $N$ gauge theory interpretation.
We show that the  large $N$ gauge theory interpretation holds at large
momentum scale
beyond the noncommutativity scale. The noncommutative Yang-Mills theory
reduces to
ordinary Yang-Mills theory in the small momentum scale limit.
Therefore there is a crossover at the noncommutativity scale.

The organization of this paper is as follows.
In section 2, we study the correlation functions
of Wilson loops in twisted reduced models.
In section 3, we apply the results of section 2 to
the correlation functions in IIB matrix  model
with D-brane backgrounds.
In section 4, we explain that the results of Seiberg and Witten can be
simply understood in our formalism. Section 5 is devoted
to conclusions and discussions.

\section{Correlators in noncommutative Yang-Mills}
\setcounter{equation}{0}

In this section, we study the correlation functions in noncommutative
Yang-Mills.
We consider twisted reduced models as a concrete realization of
noncommutative Yang-Mills.
Reduced models are defined by the dimensional reduction of $d$
dimensional gauge theory down to zero dimension (a point)\cite{RM}.
We consider $d$ dimensional $U(n)$ gauge theory coupled to
adjoint matter as an example:
\beq
S=-\int d^dx {1\over g^2}Tr({1\over 4}[D_{\mu},D_{\nu}][D_{\mu},D_{\nu}]
+{1\over 2}\bar{\psi}\Gamma _{\mu}[D_{\mu},\psi ]) ,
\eeq
where $\psi$ is a Majorana spinor field.
The corresponding reduced model is
\beq
S=- {1\over g^2}Tr({1\over4}[A_{\mu},A_{\nu}][A_{\mu},A_{\nu}]
+{1\over 2}\bar{\psi}\Gamma _{\mu}[A_{\mu},\psi ]) .
\eeq
Now $A_\mu$ and $\psi$ are $n\times n$ Hermitian matrices
and each component of $\psi$ is $d$-dimensional Majorana-spinor.
We expand the theory around the following classical solution:
\beq
[\hat{p}_{\mu},\hat{p}_{\nu}]=iB_{\mu\nu} ,
\eeq
where  $B_{\mu\nu}$ are $c$-numbers.
We assume the rank of $B_{\mu\nu}$
to be $\tilde{d}$ and define its inverse $C^{\mu\nu}$ in $\tilde{d}$
dimensional subspace.
The directions orthogonal to the subspace
is called the transverse directions.
$\hat{p}_{\mu}$ satisfy the canonical commutation relations and
they span the $\tilde{d}$ dimensional phase space.
The semiclassical correspondence shows that the
volume of the phase space is $V_p=n(2\pi)^{\tilde{d}/2} \sqrt{detB}$.

We expand $A_{\mu}=\hat{p}_{\mu}+\hat{a}_{\mu}$. We Fourier decompose
$\hat{a}_{\mu}$ and $\hat{\psi}$ fields as
\beqa
\hat{a}&=&\sum_k \tilde{a}(k) exp(iC^{\mu\nu}k_{\mu}\hat{p}_{\nu}) ,\n
\hat{\psi}&=&\sum_k \tilde{\psi}(k) exp(iC^{\mu\nu}k_{\mu}\hat{p}_{\nu}) ,
\label{twist}
\eeqa
where $exp(iC^{\mu\nu}k_{\mu}\hat{p}_{\nu})$ is the eigenstate of
adjoint $P_{\mu}=[\hat{p}_{\mu},~]$ with the eigenvalue $k_{\mu}$.
The Hermiticity requires that $\tilde{a}^* (k)=\tilde{a}(-k)$ and
$\tilde{\psi} ^*(k)=\tilde{\psi} (-k)$.
Let us consider the case that $\hat{p}_{\mu}$ consist of $\tilde{d}/2$
canonical pairs $(\hat{p}_i,\hat{q}_i)$ which satisfy
$[\hat{p}_i,\hat{q}_j]=iB\delta_{ij}$.
We also assume that the  solutions possess the discrete symmetry
which exchanges canonical pairs and $(\hat{p}_i\leftrightarrow \hat{q}_i)$
in each canonical pair.
We then find $V_{p}=\Lambda^{\tilde{d}}$ where $\Lambda$ is the extension of
each
$\hat{p}_{\mu}$.
The volume of the unit quantum in phase space is
$\Lambda^{\tilde{d}}/n=\lambda^{\tilde{d}}$
where $\lambda$ is the spacing of the quanta.
$B$ is related to $\lambda$ as $B=\lambda^2/(2\pi)$.
Let us assume the topology of the world sheet to be
$T^{\tilde{d}}$ in order to determine the distributions of $k_{\mu}$.
Then we can formally construct $\hat{p}_{\mu}$ through unitary matrices
as $\gamma_{\mu}=exp(i2\pi\hat{p}_{\mu}/\Lambda)$.
The polynomials of $\gamma_{\mu}$ are the basis of
$exp(iC^{\mu\nu}k_{\mu}\hat{p}_{\nu})$. We can therefore assume that
$k_{\mu}$ is
quantized in the unit of $|k^{min}|=\lambda/n^{1/ \tilde{d}}$.
The eigenvalues of $\hat{p}_{\mu}$ are quantized in the unit of
$\Lambda/n^{2/ \tilde{d}}=\lambda/n^{1/ \tilde{d}}$.
Hence we restrict the range of $k_{\mu}$ as
$-n^{1/ \tilde{d}} \lambda/2<k_{\mu}< n^{1/ \tilde{d}}\lambda/2$.
So $\sum_k$ runs over $n^2$ degrees
of freedom which coincide with those of $n$ dimensional Hermitian matrices.

We can construct a map from a matrix to a function
as
\beq
\hat{a} \rightarrow a(x)=\sum_k \tilde{a}(k) exp(ik\cdot x) ,
\label{proj}
\eeq
where $k\cdot x=k_{\mu}x^{\mu}$.
By this construction, we obtain the $\star$ product
\beqa
\hat{a}\hat{b} &\rightarrow& a(x)\star b(x),\n
a(x)\star b(x)&\equiv&exp({C^{\mu\nu}\over 2i}{\partial ^2\over
\partial\xi^{\mu}
\partial\eta^{\nu}})
a(x+\xi )b(x+\eta )|_{\xi=\eta=0} .
\label{star}
\eeqa
The operation $Tr$ over matrices can be exactly mapped onto the integration
over functions as
\beq
Tr[\hat{a}] =
\sqrt{det B}({1\over 2\pi})^{\tilde{d}\over 2}\int d^{\tilde{d}}x a(x) .
\label{traceint}
\eeq
The twisted reduced model can be shown to be equivalent to
noncommutative Yang-Mills by the
the following map from matrices onto functions
\footnote{A Similar mapping which corresponds to a different ordering has  been
considered  in the field theory of the lowest Landau level
fermions\cite{IKS}.}:
\beqa
\hat{a} &\rightarrow& a(x) ,\n
\hat{a}\hat{b}&\rightarrow& a(x)\star b(x) ,\n
Tr&\rightarrow&
\sqrt{det B}({1\over 2\pi})^{\tilde{d}\over 2}\int d^{\tilde{d}}x .
\label{momrule}
\eeqa
The following commutator is mapped to the covariant derivative:
\beq
[\hat{p}_{\mu}+\hat{a}_{\mu},\hat{o}]\rightarrow
{1\over i}\partial_{\mu}o(x)+a_{\mu}(x)\star o(x)-o(x)\star a_{\mu}(x)
\equiv [D_{\mu},o(x)] ,
\label{pcovder}
\eeq
We may interpret  the newly emerged coordinate space
as the semiclassical limit of $\hat{x}^{\mu}=C^{\mu\nu}\hat{p}_{\nu}$.
The space-time translation is realized by the following unitary
operator:
\beq
exp(i\hat{p}\cdot d)\hat{x}^{\mu}
exp(-i\hat{p}\cdot d)\n
=\hat{x}^{\mu}+d^{\mu} .
\label{transl}
\eeq

Applying the rule eq.(\ref{momrule}), the bosonic action becomes
\beqa
&&-{1\over 4g^2}Tr[A_{\mu},A_{\nu}][A_{\mu},A_{\nu}]\n
&=&
{\tilde{d}nB^2\over 4g^2}-\sqrt{det B}({1\over 2\pi})^{\tilde{d}\over 2}
\int d^{\tilde{d}}x
{1\over g^2} ({1\over 4}
[D_{\alpha},D_{\beta}][D_{\alpha},D_{\beta}]\n
&&+{1\over 2}[D_{\alpha},\varphi_{\nu}][D_{\alpha},\varphi_{\nu}]
+{1\over 4}[\varphi_{\nu},\varphi_{\rho}]
[\varphi_{\nu},\varphi_{\rho}])_{\star} .
\eeqa
In this expression, the indices $\alpha,\beta$ run over $\tilde{d}$
dimensional world volume directions  and $\nu,\rho$
over the transverse directions.
We have replaced $a_{\nu}\rightarrow\varphi_{\nu}$ in the transverse
directions. Inside $(~~)_{\star}$, the products should be understood as
$\star$
products and hence commutators do not vanish.
The fermionic action becomes
\beqa
&&{1\over g^2}Tr\bar{\psi}{\Gamma}_{\mu}[A_{\mu},\psi]\n
&=&
\sqrt{det B}({1\over 2\pi})^{\tilde{d}\over 2}
\int d^{\tilde{d}}x{1\over g^2}
(\bar{\psi}{\Gamma}_{\alpha}[D_{\alpha},\psi ]
+\bar{\psi}\Gamma_{\nu}[\varphi_{\nu},\psi ])_{\star} .
\eeqa
We therefore find noncommutative U(1) gauge theory.

In order to obtain noncommutative Yang-Mills theory with $U(m)$ gauge group,
we consider new classical solutions which are obtained by replacing each
element
of $\hat{p}_{\mu}$ by the $m\times m$ unit matrix:
\beq
\hat{p}_{\mu} \rightarrow \hat{p}_{\mu}\otimes {\mathbf{1}}_m .
\eeq
We require $N=mn$ dimensional matrices for this construction.
The fluctuations around this background $\hat{a}$ and $\hat{\psi}$
can be Fourier
decomposed in the analogous way as in eq.(\ref{twist}) with  $m$ dimensional
matrices $\tilde{a}(k)$ and $\tilde{\psi} (k)$ which satisfy
$\tilde{a}(-k)=\tilde{a}^{\dagger}(k)$ and
$\tilde{\psi}(-k)=\tilde{\psi} ^{\dagger}(k)$.
It is then clear that $[\hat{p}_{\mu}+\hat{a}_{\mu},\hat{o}]$ can be mapped
onto the
nonabelian covariant derivative $[D_{\mu},o(x)]$ once we use $\star$ product.
Applying our rule (\ref{momrule}) to the action in this case,
we obtain
\beqa
&&{\tilde{d}NB^2\over 4g^2}-\sqrt{det B}({1\over 2\pi})^{\tilde{d}\over 2}
\int d^{\tilde{d}}x
{1\over g^2} tr({1\over 4}
[D_{\alpha},D_{\beta}][D_{\alpha},D_{\beta}]\n
&&+{1\over 2}[D_{\alpha},\varphi_{\nu}][D_{\alpha},\varphi_{\nu}]
+{1\over 4}[\varphi_{\nu},\varphi_{\rho}][\varphi_{\nu},\varphi_{\rho}]\n
&&+{1\over 2}\bar{\psi}{\Gamma}_{\alpha}[D_{\alpha},\psi ]
+{1\over 2}\bar{\psi}\Gamma_{\nu}[\varphi_{\nu},\psi ])_{\star} .
\eeqa
where $tr$ denotes taking trace over $m$ dimensional subspace.
The Yang-Mills coupling is found to be $g^2_{NC}=(2\pi )^{\tilde{d}\over
2}g^2/B^{\tilde{d}/2}$.
Therefore it will decrease if the density of quanta in phase space
decreases with fixed $g^2$.

The Hermitian models are invariant under the unitary transformation:
$A_{\mu}\rightarrow UA_{\mu}U^{\dagger},
\psi \rightarrow U\psi U^{\dagger}$. As we shall see, the gauge
symmetry can be embedded in the $U(N)$ symmetry.
We expand $U=exp(i\hat{\lambda} )$ and parameterize
\beq
\hat{\lambda}=\sum_k \tilde{\lambda} (k)
exp(i{k}\cdot\hat{x}) .
\eeq
Under the gauge transformation, we find the fluctuations around
the fixed background transform as
\beqa
\hat{a}_{\mu}&\rightarrow & \hat{a}_{\mu}+i[\hat{p}_{\mu},\hat{\lambda} ]
-i[\hat{a}_{\mu},\hat{\lambda} ] ,\n
\hat{\psi}&\rightarrow & \hat{\psi} -i[\hat{\psi},\hat{\lambda} ] .
\eeqa
We can map these transformations onto the gauge transformation
in noncommutative Yang-Mills by our rule eq.(\ref{momrule}):
\beqa
&&a_{\alpha}(x) \rightarrow a_{\alpha}(x) +
{\partial \over \partial x^{\alpha}}\lambda (x)
-ia_{\alpha}(x)\star \lambda (x)+i\lambda (x)\star a_{\alpha}(x) ,\n
&&\varphi_{\nu}(x) \rightarrow \varphi_{\nu}(x)
-i\varphi_{\nu}(x)\star \lambda (x)+i\lambda (x)\star \varphi_{\nu}(x) ,\n
&&\psi (x)\rightarrow \psi (x)
-i\psi (x)\star \lambda (x)+i\lambda (x)\star \psi (x) .
\eeqa

The equivalence of twisted reduced models
and field theory on noncommutative space-time is very generic not restricted
to gauge theory.
We consider a noncommutative
$\phi^3$ field theory as a simple example.
The matrix  model action is given by
\beq
S = Tr \left(  -{1\over 2} [\hat{p}_{\mu},\hat{\phi}]^2
+ \lambda \hat{\phi}^3 \right) .
\eeq
Following the same procedure as in eqs. (\ref{proj} - \ref{momrule}),
we can obtain a noncommutative $U(m)$ $\phi^3$ field theory:
\beq
S = \int d^{\tilde{d}}x \ tr \left(
{1\over 2} (\partial_{\mu} \phi(x))^2 + \lambda ' \phi(x)^3
\right)_{\star}.
\eeq
Here $tr$ means a trace over $(m \times m)$ matrices and
$\lambda '=\lambda ({2\pi / B})^{\tilde{d}/ 4}$.
We recall the space-time translation operator eq.(\ref{transl}).
We construct the correlation functions by Parisi prescription\cite{RM}:
\beqa
&&<Tr[\hat{\phi}(0)\hat{\phi}(y_1)\cdots
\hat{\phi}(y_l)exp(ik^1\cdot\hat{x})]\n
&\times&Tr[\hat{\phi}(0)\hat{\phi}(z_1)\cdots
\hat{\phi}(z_l)exp(ik^2\cdot\hat{x})]
\cdots > ,
\label{phicor}
\eeqa
where
\beq
\hat{\phi}(y)=exp(i\hat{p}\cdot {y})\hat{\phi}~exp(-i\hat{p}\cdot y)
=\sum_k \tilde{\phi} (k)exp(ik\cdot(\hat{x}+y)).
\label{phitl}
\eeq
Eq.(\ref{phicor}) can be shown to be equal to the following correlator in
noncommutative
$\phi^3$ field theory:
\beqa
&&<(det{B})^{1\over 2}({1\over 2\pi})^{\tilde{d}\over 2}\int d^{\tilde{d}}x_1
tr[\phi(x_1)\phi(x_1+y_1)\cdots \phi(x_1+y_l)exp(ik^1\cdot x^1)]_{\star}\n
&\times&(det{B})^{1\over 2}({1\over 2\pi})^{\tilde{d}\over 2}\int
d^{\tilde{d}}x_2
tr[\phi(x_2)\phi(x_2+y_1)\cdots \phi(x_2+y_l)exp(ik^2\cdot x^2)]_{\star}
\cdots > .
\eeqa
Here we have generalized the definition of $\star$ product as
\beq
a(x)\star b(y) \equiv
exp({C^{\mu\nu}\over 2i}{\partial ^2\over \partial x^{\mu}\partial y^{\nu}})
a(x)b(y) .
\eeq
This correspondence can be proven by plugging $\phi (y_i)$ as expressed in eq.
(\ref{phitl}) into eq.(\ref{phicor}).
In the proof, we use the following identity:
\beq
Tr[exp(ik_1\cdot\hat{x})\cdots exp(ik_{m}\cdot\hat{x})]\n
=n\delta (\sum_ik_i )
exp({i\over 2}\sum_{i<j}C_{\mu\nu}k^{\mu}_ik^{\nu}_j) .
\label{tracefm}
\eeq

The principal goal of this paper is to study correlation functions
in noncommutative Yang-Mills theory through twisted reduced models.
Since we need to respect the gauge invariance of the theory,
we consider the gauge invariant operators namely Wilson loops.
The covariant completion of the space-time translation operator
$exp(i\hat{p}\cdot d)$ is $exp(iA\cdot d)$.
We may consider a manifestly gauge invariant Wilson loop operator such as
\beqa
W(C)&=&\lim_{m\rightarrow \infty}Tr[\prod_{j=1}^m U_j],\n
U_j&=&exp(iA\cdot \Delta d_j) ,
\eeqa
where $\Delta d_j$ is the $j$-th infinitesimal line element of a contour $C$.
After the application of our results in $\phi^3$ field theory to the above, we
obtain
\beq
\int d^4x tr[Pexp(i\int_C dy\cdot a(x+y))
\tilde{V}_C]_{\star} ,
\eeq
where $\tilde{V}_C =\prod_{k=1}^m exp(i\hat{p}\cdot \Delta d_k)$
is the translation operator along the contour $C$.
It is because
\beqa
&&\lim_{m\rightarrow \infty} \prod_{j=1}^{m}
exp(i(\hat{p}+\hat{a})\cdot \Delta d_j)\n
&=&
\lim_{m\rightarrow \infty} \prod_{j=1}^{m}
exp(i\hat{p}\cdot \Delta d_j)
exp(i\hat{a}\cdot \Delta d_j)\n
&=&
\lim_{m\rightarrow \infty} \prod_{j=1}^{m}
\tilde{V}_j
exp(i\hat{a}\cdot \Delta d_j)
\tilde{V}_j^{\dagger}\times \tilde{V}_C ,
\eeqa
where $\tilde{V}_j=\prod_{k=1}^jexp(i\hat{p}\cdot \Delta d_k)$.
$\tilde{V}_C$ carries the total momentum $k_{\mu}=B_{\nu\mu}d^{\nu}$
where $d=\sum_i\Delta d_i$ is the vector which
connects the both ends of the contour.

The total momentum $k_{\mu}$ is smaller or larger
than $\lambda$ depending
whether $d_{\mu}$ is smaller or larger than the noncommutative
length scale. When the momentum scale is much smaller than
$\lambda$, the contour is effectively closed.
These operators reduce to the gauge invariant operators in ordinary
Yang-Mills theory
by replacing $\star$ products by ordinary products.
Since $\star$ product involves derivatives,
it is legitimate to do so in the small momentum regime of the noncommutative
Yang-Mills.  The situation is very different at the momentum scale much
larger than
$\lambda$. In this case, our Wilson loop operators represent
open string like extended objects.
It must be clear by now how to construct gauge invariant
operators in noncommutative Yang-Mills theory by using twisted
reduced models.

Can we understand the high energy behavior of these correlators?
The answer is that it is identical to the large $N$ limit of
ordinary Yang-Mills theory in a sense that only planar diagrams
dominate  the perturbative summation.
Here we cite the remarkable theorem
in twisted reduced models\cite{twisted}.
It has been shown that the noncommutative phases cancel out in planar
diagrams while
they do not in nonplanar diagrams.
\par
Here we briefly explain this theorem.
We consider a noncommutative
$\phi^3$ field theory again.
Since  $x$-integral of a $\star$-product of two functions
coincides with that of an ordinary product, the propagator
becomes the ordinary one;
$\langle \tilde{\phi}(k) \tilde{\phi}(-k) \rangle \sim 1/k^2. $
On the other hand, due to the property of (\ref{tracefm}),
the $\phi^3$ vertex becomes
\beqa
Tr \hat{\phi}^3 &=& \sum_{k_1, k_2, k_3}
tr (\tilde{\phi(k_1)} \tilde{\phi(k_2)} \tilde{\phi(k_3)} )
Tr ( exp(ik_1\cdot\hat{x}) exp(ik_2\cdot\hat{x}) exp(ik_1\cdot\hat{x}) ) \n
&=& \sum_{k_1, k_2, k_3}
tr (\tilde{\phi}(k_1) \tilde{\phi}(k_2) \tilde{\phi}(k_3) ) \
n \delta (\sum_ik_i )
exp({i\over 2}\sum_{i<j}C_{\mu\nu}k^{\mu}_ik^{\nu}_j) .
\label{3vertex}
\eeqa
Thus we obtain an extra phase at each vertex
when we compare the Feynman amplitudes with those in the corresponding
commutative
theory. Note that this phase is dependent on the ordering of the matrices.
If we rewrite the incoming momenta of the three point vertex
(\ref{3vertex}) as
\beq
k_1 = l_1 - l_2, \ \ k_2 = l_2 - l_3, \ \ k_3 = l_3 - l_1,
\eeq
the phase factor in (\ref{3vertex}) can be given by
\beq
exp({i\over 2}\sum_{i<j}C_{\mu\nu}k^{\mu}_ik^{\nu}_j)
= exp({i\over 2} C_{\mu\nu} (l_1^{\mu} l_2^{\nu}
 +l_2^{\mu} l_3^{\nu}  + l_3^{\mu} l_1^{\nu}) ).
\label{vertex}
\eeq
Analogous expressions hold for generic multipoint vertices
in noncommutative field theory.
Let us adopt the double line notation for propagators
and assign a momentum  $l_i$ to each single line.
Each factor in (\ref{vertex}) $exp({i\over 2} C_{\mu\nu} l_i^{\mu} l_j^{\nu})$
can be assigned to an adjacent propagator with a momentum $l_i-l_j$
if there is any.
The total phase of a graph can be calculated by summing
the phases associated with the both ends of the propagators in this way.
For planar diagrams, the phases attached to the two ends of each propagator
cancel and the total phase of a graph is trivial
if there are no external lines. If there are external lines,
analogous factors to eq.(\ref{vertex}) remain
which involve only fixed external momenta.
For nonplanar diagrams, we must twist some propagators  or
equivalently, change the ordering of the matrices at some vertices.
Hence at twisted propagators, the phases no longer cancel.
\par
If the relevant momentum scale exceeds the noncommutativity
scale, the noncommutative phase oscillates rapidly
under the momentum integration and it kills the nonplanar
contributions. What is the coupling constant in the large $N$ gauge theory
in the high energy regime of noncommutative Yang-Mills?
Each loop contribution in the twisted reduced model can be
estimates as
\beq
{g^2\over n}m\sum_k f
=g^2m ({2\pi\over B})^{\tilde{d}\over 2}\int^{\Lambda}
{d^{\tilde{d}}k\over (2\pi)^{\tilde{d}}}f ,
\eeq
where $f$ is a Feynman integrand.
We recall that $|k^{min}|=\lambda/n^{1/\tilde{d}}$
and the integration range is
$-n^{1/\tilde{d}}\lambda /2< k_{\mu} < n^{1/\tilde{d}}\lambda /2$.
We therefore find that the 't Hooft coupling of noncommutative Yang-Mills
$g^2_{NC}m=g^2 m({2\pi /B})^{\tilde{d}/ 2}$
is also the 't Hooft coupling in the high energy large $N$ gauge
theory. Our findings here has a very important implication for the
renormalizability
of noncommutative Yang-Mills.
Our conclusion is that noncommutative Yang-Mills theory
is renormalizable if and only if the corresponding large $N$ gauge theory
is renormalizable.

We extend these constructions in continuum theory to lattice gauge theory.
We consider twisted reduced  models in pure lattice gauge theory.
The Wilson action is
\beq
-{1\over 2\tilde{g}^2}
\sum_{\mu,\nu}Tr[Z_{\mu\nu}U_{\mu}U_{\nu}U^{\dagger}_{\mu}U^{\dagger}_{\nu}] ,
\label{wilsonac}
\eeq
where $\tilde{g}$ is the lattice coupling constant.
The 't Hooft matrices are the classical solution of the model
which satisfy
\beq
\gamma_{\mu}\gamma_{\nu}=\gamma_{\nu}\gamma_{\mu}exp({2\pi i}n_{\nu\mu}) ,
\eeq
where $Z_{\mu\nu}=exp({2\pi in_{\mu\nu}})$.
A four dimensional example is
\beq
n_{\mu\nu}=\left( \begin{array}{cccc}
0 & -{1\over \sqrt{n}} & 0 & 0 \\
{1\over \sqrt{n}} & 0 & 0 & 0 \\
0 & 0 & 0 & -{1\over \sqrt{n}} \\
0 & 0 & {1\over \sqrt{n}} & 0
\end{array} \right)_.
\eeq
where we assume $\sqrt{n}$ is an integer.

In order to make contact with continuum theory,
we may interpret  $\gamma_{\mu} =
exp(iC^{\nu\mu}k^{min}_{\nu}\hat{p}_{\mu})$ where
the index
$\mu$ is not summed.
In the weak coupling regime, we parameterize
$U_{\mu}$ as
\beqa
U_{\mu}&=&exp(iC^{\nu\mu}k^{\min}_{\nu}(\hat{p}_{\mu}+\hat{a}_{\mu}))\n
&=&exp(i(\hat{p}_{\mu}+\hat{a}_{\mu})e^{\mu}) ,
\label{uparam}
\eeqa
where $|e^{\mu}|=2\pi/(\lambda n^{1/ \tilde{d}})$.
Since $\gamma_{\mu}=exp(i\hat{p}_{\mu}e^{\mu})$ is the translation operator,
we may regard $U_{\mu}$ as a link variable on the square lattice with the
lattice spacing $|e^{\mu}|$.
$\gamma_{\mu}$ can be reexpressed as $exp(ik^{\min}_{\nu}\hat{x}^{\nu})$
where $\hat{x}^{\nu}=C^{\nu\mu}\hat{p}_{\mu}$ and $\mu$ should not be
summed in this paragraph.
It also implies that the target space is
the ${\tilde{d}}$ dimensional noncommutative torus with the radius
$n^{1/\tilde{d}}/\lambda$.
We need to take large $n$ limit to let the lattice spacing vanish and
obtain continuum limit.  Simultaneously we obtain infinitely extended
space-time
since the radius of the torus diverges in this limit.

We Fourier decompose $\hat{a}_{\mu}$ as
\beq
\hat{a}=\sum_{k}a(k)exp(ik\cdot \hat{x}) ,
\eeq
where
$k_{\mu}=n_{\mu}k^{\min}_{\mu}$ and $n_{\mu}$ are integer valued.
The Hermiticity requires that $a(k)^{\dagger}=a(-k)$
and we sum over $n^2$ degrees of freedom.
$exp(ik\cdot \hat{x})$ can be expressed by the 't Hooft matrices as
\beq
exp(ik\cdot \hat{x})=\gamma_{0}^{n_0}\gamma_{1}^{n_1}\cdots
\gamma_{\tilde{d}-1}^{n_{\tilde{d}-1}}\gamma_{\tilde{d}}^{n_{\tilde{d}}}
exp({\pi i}<n|n>) .
\eeq
where $<k|q>=\sum_{\mu <\nu} k_{\mu}q_{\nu}n_{\nu\mu}$.
Eq. (\ref{tracefm}) holds exactly in lattice formulation.

By using the parameterization eq.(\ref{uparam}), the Wilson action becomes
\beq
-{\tilde{d}(\tilde{d}-1)nm\over 2\tilde{g}^2}-{({2\pi})^4\over
4\tilde{g}^2n^{4\over \tilde{d}}
\lambda^4}Tr([p_{\mu}+a_{\mu},p_{\nu}+a_{\nu}]-iB_{\mu\nu})^2 +\cdots .
\eeq
The mapping rule eq.(\ref{momrule}) can be rigorously justified within
lattice gauge
theory
where the topology of the world volume is $T^{\tilde{d}}$.
We can now apply our rule to obtain noncommutative
Yang-Mills in a naive continuum limit.
\beq
-{\tilde{d}(\tilde{d}-1)nm\over 2\tilde{g}^2}-{1\over 4\tilde{g}^2n
|e^{\mu}|^{\tilde{d}-4}}\int
d^{\tilde{d}}x tr([D_{\mu},D_{\nu}][D_{\mu},D_{\nu}])_{\star}
+\cdots .
\eeq

Here we find the relationship between the coupling of the twisted reduced model
on the lattice and the coupling of the noncommutative Yang-Mills theory
$g^2_{NC}$:
\beq
g^2_{NC}m=\tilde{g}^2nm
|e^{\mu}|^{\tilde{d}-4} .
\eeq
We find again that the 't Hooft coupling of noncommutative Yang-Mills
is identical to the 't Hooft coupling of high energy large $N$ gauge theory.
Since the lattice spacing $|e^{\mu}|$ depends on $n$ for fixed $\lambda$,
$n$ dependence of $g^2_{NC}$ becomes nontrivial in general.

An example of the Wilson loop is
\beq
W(C)=Tr[
\prod_{i \in C} {U}_i] ,
\label{naivewil}
\eeq
where ${U}_i$ denotes link variables
along a contour $C$.
Let us parameterize ${U}_j=exp(i(\hat{p}+\hat{a})\cdot e_j)$.
We then define $V_j=exp(i\hat{p}\cdot e_j)$.
The Wilson loop can be expressed as
\beq
Tr[\prod_{j \in C}{U}(x_j)\times\tilde{V}_C] ,
\label{Wilsloop}
\eeq
where $\tilde{V}_C=P\prod_{j\in C}V_j$.
$\tilde{V}_C$ is the translation operator along the Wilson loop on the lattice.
${U}(x_j)$ is defined as
\beqa
{U}(x_j)&=&\tilde{V}_jV^{\dagger}_j{U}_j\tilde{V}_j^{\dagger}\n
&=&\tilde{V}_jexp(-i\hat{p}\cdot e_j)
exp(i(\hat{p}+\hat{a})\cdot e_j)\tilde{V}_j^{\dagger}\n
&=&\tilde{V}_jexp(i\hat{a}\cdot e_j
+{1\over 2}[\hat{p}\cdot e_j,\hat{a}\cdot e_j]+\cdots )
\tilde{V}_j^{\dagger} ,
\eeqa
where $\tilde{V}_j=V_1V_2\cdots V_j$.
Since $\tilde{V}_j$ is the translation operator along the contour $C$,
${U}(x_j)$
can be regarded as the link variable at the $j$-th link on $C$.
We may associate the Wilson loop eq.(\ref{Wilsloop}) with the total momentum
$k_{\mu}=\sum_j k^j_{\mu}$ where $V_j=exp(ik^j\cdot\hat{x})$.

In the conventional interpretation,
the gauge invariant Wilson loops of gauge theory are identified with those with
vanishing total momenta in reduced models.
In this case $\tilde{V}_C$ reduces to a pure phase.
It has been shown that the noncommutative phase factors cancel
in planar diagrams while nontrivial phases remain in nonplanar diagrams.
Based on this fact, it was argued that
the correlation functions of the Wilson loops $Tr[\prod_{j \in C}{U}(x_j)]$
in twisted reduced models and in the large $N$ limit of gauge theory are
identical\cite{twisted}
since the contributions from nonplanar diagrams vanish due to the rapid
oscillations of the phases. In view of our noncommutative Yang-Mills theory
interpretation of the twisted reduced models, we now believe that this
argument needs
to be reconsidered.
Namely the phase factors reduce to identity if the relevant momentum scale is
smaller than the noncommutative scale $\lambda$. Although the expectation
value of
the  Wilson loops with length scale less than $1/\lambda$ agrees with the
large $N$
limit of gauge theory, there is a crossover at this scale and it is
eventually described by $U(m)$ gauge theory at long distances!

The major advantage of the noncommutative Yang-Mills interpretation
of the twisted reduced model is that we can interpret
the Wilson loops eq.(\ref{Wilsloop}) with nonvanishing total momenta $k_{\mu}
=n_{\mu}k^{min}_{\mu}$.
We have argued heuristically that
$n_{\mu}$ may be interpreted as the winding numbers of the Wilson loops in
large $N$ gauge
theory with compactification radius $|e^{\mu}|$. If reduced models are
string theory,
the winding modes should be able to be interpreted as the momenta due to
T duality.
We have firmly established the validity of these arguments by showing that
$k_{\mu}
$ which is smaller than $\lambda$
can be interpreted as a momentum. In this process we have found that
noncommutative
Yang-Mills theory appears in our dual interpretation of large $N$ gauge theory.
The Wilson loops with $k_{\mu}$ which is much larger than $\lambda$
represent extended objects in the continuum limit.

We have pursued the possibility that a version of reduced model
defines string theory. We may postulate that the simple bosonic twisted
reduced model
eq.(\ref{wilsonac}) is a string theory in the spirit of
\cite{Zacos}\cite{Bars}.
Let us examine this postulate in view of our findings in this section.
This statement makes sense in four dimensions since the large $N$ gauge theory
is believed to be confining. In fact we can estimate the string tension
by using the equivalence of twisted
reduced models and large $N$ gauge theory at high energy regime:
\beq
\alpha '\sim {1\over \Lambda^2}exp({1\over b_0\tilde{g}^2N}) ,
\label{tension}
\eeq
where $b_0=1/(4\pi )^2(11/3)$ and $\Lambda^2 \sim B\sqrt{n}$.
In order to achieve universality, we need to approach the weak coupling
limit by letting the 't Hooft coupling $\tilde{g}^2N \rightarrow 0$
in such a way that $exp({1/ b_0\tilde{g}^2N}) \sim {n}^{\alpha}$ with positive
definite exponent $\alpha$.
The exponent $\alpha$ must be smaller than $1/2$ for this estimate to be valid.
If not, the coupling constant remains small at the noncommutativity scale
and we obtain $U(m)$ gauge theory at low energy regime.

\section{Correlators on D-branes in IIB matrix model}
\setcounter{equation}{0}

In this section we apply the results of the preceding section
to the correlation functions in IIB matrix model with
D-brane backgrounds.
We need to interpret $A_{\mu}$ as coordinates in IIB matrix model
due to $\cal{N}$=2 SUSY as we have emphasized in the introduction.
For this purpose, we identify the solution of IIB matrix model as
$\hat{x}^{\mu}$ which satisfy.
\beq
[\hat{x}^{\mu},\hat{x}^{\nu}]=-iC^{\mu\nu} .
\label{background}
\eeq
Now the plane waves correspond to
the eigenstates of $\hat{P}_{\mu}=[\hat{p}_{\mu},~]$ with small eigenvalues,
where
$\hat{p}_{\mu}=B_{\mu\nu}\hat{x}^{\nu}$.
$\hat{x}^{\mu}$ and $\hat{p}_{\nu}$ satisfy the canonical commutation
relation: $[\hat{x}^{\mu},\hat{p}_{\nu}]=i\delta^{\mu}_{\nu}$.
We expand $A_{\mu}=\hat{x}_{\mu}+\hat{a}_{\mu}$ as before and
$\hat{a}_{\mu}$ and
$\hat{\psi}$ can be Fourier decomposed as in
\beqa
\hat{a}&=&\sum_k \tilde{a}(k) exp(i{k}\cdot\hat{x}),\n
\hat{\psi}&=&\sum_k \tilde{\psi}(k) exp(i{k}\cdot\hat{x}).
\label{twistbr}
\eeqa

Once the eigenvalues of $[\hat{A}_{\mu},~]$ are identified with momenta,
the coordinate space has to be embedded in the rotated matrices
$C^{\mu\nu}\hat{p}_{\nu}$
as we have seen in section 2.
If we identify the large eigenvalues of $A^{\mu}$ as the coordinates
$x^{\mu}$,
we have to rotate the covariant derivatives as follows:
\beq
[\hat{x}^{\mu}+\hat{a}^{\mu},\hat{o}]
\rightarrow C^{\mu\nu}({1\over i}\partial_{\nu}o(x)+b_{\nu}(x)\star o(x)
-o(x)\star b_{\nu}(x)).
\label{covder}
\eeq
Note that we have defined a new gauge field $b_{\mu}(x)$ by this expression.
We can map the matrices onto functions by using
the rule eq.(\ref{momrule}).
The relevance of $\star$ product to D-string action in IIB matrix
model was noted by Li\cite{Li}.

We study D3 brane backgrounds in this section.
A D3-brane solution may be constructed as follows:
\beqa
A_0 &=& \frac{T}{\sqrt{2\pi n_1}}\hat{q} \equiv \hat{x}^0 ,\n
A_1 &=& \frac{L}{\sqrt{2\pi n_1}}\hat{p} \equiv \hat{x}^1 ,\n
A_2 &=& \frac{L}{\sqrt{2\pi n_2}}\hat{q}' \equiv \hat{x}^2 ,\n
A_3 &=& \frac{L}{\sqrt{2\pi n_2}}\hat{p}' \equiv \hat{x}^3 ,\n
\mbox{other }A_{\mu} \mbox{'s}&=& 0,
\eeqa
which may be embedded into $n_1n_2$ dimensional matrices.
We can further consider $m$ parallel D3-branes after  replacing each
element
of the D3-brane solution by $m \times m$ unit matrix.
Under the replacements $[A_{\alpha},\hat{o}]\rightarrow
C^{\alpha\beta}[D_{\beta},o(x)]$ and
$A_{\nu}\rightarrow (1/B)\varphi_{\nu}$,
the bosonic action becomes
\beqa
&&-{1\over 4g^2}Tr[A_{\mu},A_{\nu}][A_{\mu},A_{\nu}]\n
&=&{mTL^3\over (2\pi )^2 g^2}
-{1\over B^2(2\pi )^2g^2}\int d^4x tr({1\over 4}[D_{\alpha},D_{\beta}]
[D_{\alpha},D_{\beta}]\n
&&+{1\over 2}[D_{\alpha},\varphi_{\nu}][D_{\alpha},\varphi_{\nu}]
+{1\over 4}[\varphi_{\nu},\varphi_{\rho}]
[\varphi_{\nu},\varphi_{\rho}])_{\star} ,
\eeqa
where we integrate over the four dimensional world volume of
D3-branes.
As for the fermionic action, we find
\beqa
&&Tr\bar{\psi}\Gamma_{\mu}[A_{\mu},\psi]\n
&=&
\int d^4x tr(\bar{\psi}\tilde{\Gamma}_{\alpha}[D_{\alpha},\psi ]
+\bar{\psi}\Gamma_{\nu}[\varphi_{\nu},\psi ])_{\star} .
\eeqa
We thus find four dimensional $\cal{N}$=4 super Yang-Mills theory.
The coupling of noncommutative Yang-Mills is found to be
$g^2_{NC}=g^2B^2(2\pi )^2$.
Recall that $(2\pi /B)^2 =R^4$ is the unit volume of a quantum
and $R$ is the average spacing.
The coupling $g^2_{NC}$ is a function of eigenvalue density
of the matrices.
The low and high density regimes
correspond to weak and strong coupling regimes of
noncommutative Yang-Mills respectively.

We have proposed that the fundamental strings are created by the
Wilson loop operators. Simple examples are the following vertex operators
for a dilaton, an axion and gravitons
which are consistent with the interactions of D-instantons\cite{AIIKKT}:
\beqa
&&Tr\{([A_{\alpha},A_{\beta}]+iC_{\alpha\beta})^2
exp(ik\cdot A)\}
+ \mbox{fermionic terms} ,\n
&&\epsilon_{\alpha\beta\gamma\delta}
Tr\{([A_{\alpha},A_{\beta}]+iC_{\alpha\beta})
([A_{\gamma},A_{\delta}]+iC_{\gamma\delta})
exp(ik\cdot A)\}
+ \mbox{fermionic terms} ,\n
&&Tr\{([A_{\alpha},A_{\mu}]+iC_{\alpha\mu})
([A_{\mu},A_{\beta}]+iC_{\mu\beta})
exp(ik\cdot A)\}
+ \mbox{fermionic terms} .
\label{highver}
\eeqa
We have the corresponding vertex operators in CFT\cite{maldacena}:
\beqa
&&\int d^4xtr([D_{\alpha},D_{\beta}][D_{\alpha},D_{\beta}])exp(ik\cdot x)
+ \mbox{fermionic terms},\n
&&\int d^4x\epsilon_{\alpha\beta\gamma\delta}
tr([D_{\alpha},D_{\beta}][D_{\gamma},D_{\delta}])exp(ik\cdot x)
+ \mbox{fermionic terms},\n
&&\int d^4xtr([D_{\alpha},D_{\gamma}][D_{\gamma},D_{\beta}]+
[D_{\alpha},\varphi_{\nu}][D_{\beta},\varphi_{\nu}])exp(ik\cdot x)
+ \mbox{fermionic terms} .
\label{dagop}
\eeqa

We now apply our rule eq.(\ref{momrule}) to the Wilson loops
with the replacements $[A_{\alpha},\hat{o}]\rightarrow C^{\alpha\beta}
[D_{\beta},o(x)],A_{\nu}\rightarrow 1/B\varphi_{\nu}$.
The Wilson loops in IIB matrix model can be mapped onto those in
noncommutative Yang-Mills as it is explained in the previous section.
For example,
\beqa
&&Tr\{([A_{\alpha},A_{\mu}]+iC_{\alpha\mu})
([A_{\mu},A_{\beta}]+iC_{\mu\beta})
exp(ik\cdot A)\}\n
&\rightarrow&
\int d^4 x tr\left( ([D_{\alpha},D_{\gamma}][D_{\gamma},D_{\beta}]+
[D_{\alpha},\varphi_{\nu}][D_{\beta},\varphi_{\nu}])
Pexp(i\int _Cdy\cdot b(x+y))exp(ik\cdot x)\right)_{\star} .\n
\label{lowver}
\eeqa
where $C$ is a straight path with the length $|C^{\mu\nu}k_{\nu}|$.
They reduce to the vertex operators in ordinary gauge theory
in low energy regime.
On the other hand,
they represent open string like extended objects
in the high energy regime.

In noncommutative space-time, it is not possible to consider
states which are localized
in the domain whose volume is smaller than the noncommutative scale.
Therefore if we consider the states with
large momentum or small longitudinal length scale,
they must expand in the transverse directions.
Recall that the both momentum space and coordinate
space are embedded in the matrices of twisted reduced models.
They are related by $\hat{x}^{\mu}=C^{\mu\nu}\hat{p}_{\nu}$.
The momenta $k_{\mu}$ are the eigenvalues of adjoint
$\hat{P}_{\mu}=[\hat{p}_{\mu},~]$. The corresponding eigenstates
such as $exp(ik^1\cdot\hat{x})$ and $exp(ik^2\cdot\hat{x})$
are not commutative to each other if $|k^i_\mu| > \lambda$
since $exp(ik^1\cdot\hat{x})exp(ik^2\cdot\hat{x})
=exp(ik^2\cdot\hat{x})exp(ik^1\cdot\hat{x})exp(iC^{\mu\nu}k^1_{\mu}k^2_{\nu})$.
They may be interpreted as string like extended objects whose
length is $|C^{\mu\nu}k_{\nu}|$.
The vertex operator eq.(\ref{lowver}) appears to be consistent with such an
interpretation.
In our interpretation, the Wilson loop
expands in the orthogonal direction to the momentum which is consistent with
the commutation relations $[\hat{x}^{\mu},\hat{p}_{\nu}]=i\delta^{\mu}_{\nu}$
and $[\hat{x}^{\mu},\hat{x}^{\nu}]=-iC^{\mu\nu}$.

As we have argued, there is a crossover at the momentum scale $\lambda$.
When we consider the Wilson loops, we find that
the planar diagrams dominate at larger momentum scale than $\lambda$ and the
diagrams
of all topology contribute in lower momentum scale than $\lambda$.
It may be interpreted as that the string coupling (dilaton expectation value)
is scale dependent.
It is because in our IIB matrix model conjecture,
the tree level string theory is considered to be obtained by
summing planar diagrams and string perturbation theory
is identified with the topological expansion of the matrix model.
With this interpretation, the string coupling
grows as the relevant momentum scale is decreased while it vanishes
in the opposite limit. In the $D$ brane interpretation, the small momentum
region
corresponds to the vicinity of the brane, while the large momentum region
corresponds to
the region far from the brane since the Higgs expectation value plays the
same role with the
momentum scale.

The dilaton expectation value in $AdS_5\times S_5$ with constant NS $B$ field
background $b$ is\cite{ads+f}\cite{hashimoto}
\beq
e^{2\phi}=g_s^2{(1+{\alpha'^2g_sm(1+b^2)\over r^4})^2
\over (1+{\alpha'^2g_sm\over r^4})^2} ,
\eeq
where $r$ is the distance from the D3-brane.
The metric is
\beq
ds^2=(1+{\alpha'^2g_sm(1+b^2)\over r^4})^{1\over 2}(
{d\vec{x}^2\over 1+{\alpha'^2g_sm\over r^4}}+
dr^2 +r^2d\Omega_5^2 ) .
\eeq
We consider large $b$ and small $g_s$ limit while keeping $g_sb^2$ fixed:
\beqa
e^{2\phi}&\rightarrow&
g_s^2(1+{\alpha'^2g_smb^2\over r^4})^2 ,\n
ds^2&\rightarrow&
(1+{\alpha'^2g_smb^2\over r^4})^{1\over 2}(
{d\vec{x}^2}+
dr^2 +r^2d\Omega_5^2 ) .
\label{adslr}
\eeqa
$e^{2\phi}$ behaves just like the twisted reduced models as we just explained
in the preceding paragraph.
The supergravity description is valid when $g_smb^2$ is large.
As it will be explained in the subsequent section
after eq.(\ref{gseqn}),
the coupling of noncommutative
Yang-Mills is given by $g_s b^2$ when $b$ is large.
It sets the radius of $AdS_5$ and $S_5$ as well.
Therefore supergravity description is valid in the strong
coupling limit of noncommutative Yang-Mills.

As we have shown in \cite{AIIKKT}, IIB matrix model is capable
to describe flat space-time which is realized at large $r$ region in
eq.(\ref{adslr}) which we believe is an advantage over $AdS$/CFT.
The crossover region of noncommutative Yang-Mills
may be understood by supergravity
since in this region ($g_sm << r^4/\alpha'^2 <<g_smb^2$)
\beq
ds^2\rightarrow
({\alpha'^2g_smb^2\over r^4})^{1\over 2}(
{d\vec{x}^2}+
dr^2 +r^2d\Omega_5^2 ) .
\eeq
It is amusing to note that the branes literally exist
at the boundary of $AdS_5$ in this interpretation.
Another limit we may consider is large $m$ limit
keeping $g_sm$ fixed and large:
\beqa
e^{2\phi}&\rightarrow &g^2_s (1+b^2)^2,\n
ds^2&\rightarrow&({\alpha'^2g_sm(1+b^2)\over r^4})^{1\over 2}(
{d\vec{x}^2\over {\alpha'^2g_sm\over r^4}}+
dr^2 +r^2d\Omega_5^2 ) .
\eeqa
It may correspond to the $AdS$ description of low energy $U(m)$ gauge theory
of D3-branes.

What we have found here is that the crossover behavior
of the correlators in
IIB matrix model with D3-brane backgrounds
is consistent with supergravity description
which is valid in the strong coupling regime of noncommutative Yang-Mills.
Since the relevant coupling goes like $g^2/R^4$,
the strong coupling regime corresponds to high density regime in IIB matrix
model. In this regime we therefore find very weak gravity
since the theory is well described by classical supergravity.
We have also studied the opposite limit in\cite{AIKKT}.
In that regime we have found very strong gravity. The space-time appears
to be fractal instead of very flat $AdS_5\times S_5$.
It is very encouraging to find that IIB matrix model is capable to describe
both strong and weak coupling limits of gravity.
It raises the hope that it may be able to describe realistic space-time
at intermediate coupling.

\section{Correspondence with Seiberg-Witten Formulation}

We have put forward the IIB matrix conjecture that the simple
matrix model defined by the action eq.(\ref{action}) is a nonperturbative
formulation of superstring theory.
Matrix models naturally incorporate the notion of minimum
length scale by the uncertainty principle\cite{yoneya}.
We denote the spacing of the eigenvalues of matrices by $R$.
Although it is attractive to assume that there are finite quanta per
each volume element of string scale by assuming $\alpha '\sim R^2$,
we also need an entirely different scale to ensure universality.
We have seen that the large momentum cutoff $\Lambda$ is much larger than
$\lambda\sim 1/R$ as $\Lambda\sim n^{1/\tilde{d}}\lambda$ with D-brane
backgrounds.
We have also shown that the renormalizability of noncommutative Yang-Mills
is identical to
large $N$ gauge theory. We can therefore obtain universal results if we expand
the IIB matrix model action around $D3$ branes for example.
We can consider the Wilsonian effective action at string scale by
integrating out heavier degrees of freedom.
If IIB matrix model conjecture is correct, it must agree with Born-Infeld
type effective
action of string theory.

In a recent paper\cite{SW}, open string theory with
Neumann boundary condition was studied in a constant $b_{\mu\nu }$ background
\footnote{Here $b_{\mu\nu }$ denote the 2nd rank antisymmetric tensor field
in
the $(NS,NS)$-sector. We use the lower case letter to distinguish from
$B_{\mu\nu}$ in the previous sections. The correspondence between our and
their
notations is like
$b_{\mu\nu}\leftrightarrow B_{ij}, C^{\mu\nu}\leftrightarrow -\theta^{ij}$.}
,
and it was shown that one can get commutative or noncommutative description
of the theory depending on the regularization.
Chepelev and Tseytlin argued that our D-brane solutions can
be interpreted as not pure D-branes but those with nonvanishing $U(1)$ field
strength\cite{tseytlin},
while we have shown that they can be interpreted in terms
of noncommutative geometry with vanishing $U(1)$ field strength.
Since having D-branes in constant $b_{\mu\nu}$ background is gauge equivalent
to having ones with nonvanishing $U(1)$ field strength,
apparently these arguments are consistent with each other.
The background
eq.(\ref{background})
we studied can also be considered in the string theory context and
it corresponds to a D-brane with nonvanishing gauge field strength on
the worldvolume.
Therefore we suspect that
their noncommutative description and ours are physically the same.
We cannot give a direct proof of this fact but will give evidences for it
in this section.

They start from the Euclidean action
\footnote{In this section we suppress fermionic degrees of freedom
for simplicity.}
\begin{equation}
S=\frac{1}{4\pi\alpha^\prime}\int_\Sigma
(g_{\mu\nu }\partial_ax^\mu \partial^ax^\nu
-2\pi i\alpha^\prime b_{\mu\nu }\epsilon^{ab}\partial_ax^\mu
\partial_bx^\nu ),
\label{straction}
\end{equation}
where the coordinates $x^\mu ~(\mu =1\cdots ,\tilde{d})$ are along
D$p$-branes with $\tilde{d}\leq p+1$ and $\tilde{d}$ even.
If one takes Pauli-Villars regularization in treating the vertex operators
in this theory, one obtains the usual Dirac-Born-Infeld action with
the gauge field strength $F$ replaced by $F+b$ as the effective
Lagrangian for slowly varying fields:
\begin{equation}
{\cal L}
=
\frac{1}{g_s(2\pi )^p(\alpha^\prime )^{\frac{p+1}{2}}}
\sqrt{\mbox{det}(g+2\pi\alpha^\prime (b+F))}.
\label{comdes}
\end{equation}

If one takes point splitting regularization instead, one obtains a
noncommutative
description. The effective Lagrangian becomes
\begin{equation}
\hat{\cal L}
=
\frac{1}{G_s(2\pi )^p(\alpha^\prime )^{\frac{p+1}{2}}}
\left(\sqrt{\mbox{det}(G+2\pi\alpha^\prime \hat{F})}\right)_{\star},
\label{noncomdes}
\end{equation}
where all the product of the fields in this description should
be understood as the $\star$ product defined as
\begin{equation}
f(x)\star g(x)=e^{\frac{1}{2i}C^{\mu\nu }\frac{\partial}{\partial\xi^\mu }
\frac{\partial}{\partial\zeta^\nu }}
f(x+\xi )g(x+\zeta )|_{\xi =\zeta =0}.
\end{equation}
$G_{\mu\nu }$ and $C^{\mu\nu }$ in the noncommutative description are given
by
\begin{eqnarray}
G_{\mu\nu }
&=&
g_{\mu\nu }-(2\pi\alpha^\prime )^2(bg^{-1}b)_{\mu\nu },
\nonumber
\\
C^{\mu\nu }
&=&
2\pi\alpha^\prime (\frac{1}{g+2\pi\alpha^\prime b})_A^{\mu\nu },
\end{eqnarray}
where $(~)_A$ denotes the antisymmetric part of the matrix.

There exist
the interpolating descriptions between the
two above which was also proposed by Pioline and Schwarz\cite{PS}.
The effective Lagrangian is
\begin{equation}
\hat{\cal L}_\Phi
=
\frac{1}{G_s(2\pi )^p(\alpha^\prime )^{\frac{p+1}{2}}}
\left(\sqrt{\mbox{det}(G+2\pi\alpha^\prime \hat{F}+\Phi )}\right)_{\star}.
\label{Lhat}
\end{equation}
$\Phi_{\mu\nu }$ denotes an antisymmetric tensor field.
This time $G_{\mu\nu }$ and $C^{\mu\nu }$ are given as
\begin{equation}
\frac{1}{G+2\pi\alpha^\prime \Phi}
=
\frac{C}{2\pi\alpha^\prime }+\frac{1}{g+2\pi\alpha^\prime b}.
\label{phi}
\end{equation}

The D$p$-branes we are considering here can be expressed as a configuration
of
infinitely many D$(p-\tilde{d})$-branes in string theory as well as in
matrix model as follows.
A configuration of infinitely many D$(p-\tilde{d})$-branes can be expressed
by
gauge field and transverse coordinates $X^\mu$ which are in the adjoint
representation
of
$U(\infty )$.
The configuration we consider here is
\begin{equation}
X^\mu =\hat{x}^\mu ,
\label{conf}
\end{equation}
where
\begin{equation}
[\hat{x}^\mu ,\hat{x}^\nu ]=-iC^{\mu\nu}.
\end{equation}
This configuration corresponds to a D$p$-brane with gauge field strength
$C^{-1}_{\mu\nu}=B_{\mu\nu}$.
Therefore the worldvolume theory of the D$p$-brane can also be described as
the
worldvolume theory of infinitely many  D$(p-\tilde{d})$-branes.
As we saw in section 2,
the worldvolume theory of the D$p$-brane becomes noncommutative Yang-Mills
theory in this description.

Here we will show that there exists such a noncommutative description
of D$p$-brane for which
the effective action becomes eq.(\ref{Lhat}) with $G_{\mu\nu }$ and
$C^{\mu\nu }$
satisfying eq.(\ref{phi}).
In order to do so, we should consider such a configuration eq.(\ref{conf})
in
constant background $g_{\mu\nu },b^\prime_{\mu\nu }$.
The worldsheet action in the Lorentzian signature is
\begin{equation}
S_L
=
\frac{1}{4\pi\alpha^\prime}\int_\Sigma
[g_{\mu\nu }(\partial_tx^\mu \partial_tx^\nu -\partial_\sigma x^\mu
\partial_\sigma x^\nu )
+4\pi i\alpha^\prime b_{\mu\nu }^\prime\partial_tx^\mu \partial_\sigma
x^\nu ].
\label{actionB}
\end{equation}
Since the second term in the Lagrangian is in the form of a total
derivative,
it can have nontrivial effect if there are boundaries on the worldsheet.
Therefore
the open string theory with the action eq.(\ref{actionB}) and the boundary
state
$|B\rangle_{b^\prime_{\mu\nu}}$ is equivalent to the one with the action in
which
$b^\prime_{\mu\nu}=0$ and the boundary state
\begin{equation}
|B\rangle_0=e^{-\frac{i}{2}\int d\sigma
b^\prime_{\mu\nu}x^\mu\partial_\sigma x^\nu}
|B\rangle_{b^\prime_{\mu\nu}}.
\label{0B}
\end{equation}
When we consider the open string theory with Neumann boundary conditions,
the factor $e^{-\frac{i}{2}\int d\sigma b^\prime_{\mu\nu}x^\mu\partial_\sigma
x^\nu}$
 has the effect of increasing the gauge field strength
$F$ by $b^\prime$ and the physics depend only on $b^\prime +F$.

Now let us consider the open string theory corresponding to the
configuration
eq.(\ref{conf}).
The most efficient way to study open string theory corresponding to such
a configuration is to look at the boundary state.
The boundary state for such a configuration has a path integral
representation
\cite{ishibashi}
\begin{equation}
|B\rangle_{b^\prime_{\mu\nu }}=\int [d\xi ]
\exp [-\frac{i}{2}\int d\sigma \xi^\mu \partial_\sigma \xi^\nu B_{\mu\nu }
-i\int d\sigma p_\mu \xi^\mu ]|B\rangle_{inst} .
\end{equation}
$\xi$ denote the $c$-number counterpart of $\hat{x}^\mu$ in the path integral
and
the term $\int d\sigma \xi^\mu \partial_\sigma \xi^\nu B_{\mu\nu }$ gives
the symplectic form corresponding to the commutation relation
eq.(\ref{conf}).
$|B\rangle_{inst}$ is the boundary state for a D-instanton at the origin
satisfying
$x^\mu (\sigma )|B\rangle_{inst}=0$. The canonical momentum $p_\mu (\sigma )$
is now
\begin{equation}
p_\mu =\frac{1}{2\pi\alpha^\prime}(g_{\mu\nu }\partial_tx^\nu
+2\pi\alpha^\prime b^\prime_{\mu\nu }\partial_\sigma x^\nu ).
\end{equation}
 From the identity
\begin{equation}
0=\int [dP]\frac{\delta}{\delta P^{\rho}}
exp [-\frac{i}{2}\int d\sigma P^\mu \partial_\sigma P^\nu B_{\mu\nu }
-i\int d\sigma p_\mu P^\mu ]|B\rangle_{inst},
\end{equation}
we obtain an identity satisfied by
corresponding $|B\rangle_0$ as
\begin{equation}
[\frac{1}{2\pi\alpha^\prime}g_{\mu\nu }\partial_tx^\nu
+(b^\prime_{\mu\nu }+B_{\mu\nu })\partial_\sigma x^\nu ]|B\rangle_0=0.
\end{equation}
Therefore the open string theory we consider here is equivalent to
the one with Neumann boundary conditions for $x^\mu $'s but with the gauge
field
strength $F=b^\prime +B$. Thus we get D$p$-branes with $F$ or
equivalently D$p$-branes in $b=b^\prime +B$ background.

Physics does not depend on the choice of the parameter $b^\prime $ and $B$
as long as $b=b^\prime +B$ is the same. Since $B=C^{-1}$,
we have theories with different $\star$-products which are physically
equivalent.
This is the situation quite the same as the one encountered in \cite{SW}.
To compare the theory here with theirs, we should study the effective
action for slowly varying fields. In our description, what we should look at
is the effective action for the slowly varying fluctuations around the
configuration
in eq.(\ref{conf}) of the D$(p-\tilde{d})$-branes. The most convenient way
to deduce
such an action is to use T-duality. The duality transformation in all the
$x^\mu $-directions maps the action eq.(\ref{actionB}) to
\begin{equation}
\tilde{S}_L
=
\frac{1}{4\pi\alpha^\prime}\int_\Sigma
[\tilde{g}^{\mu\nu }(\partial_t\tilde{x}_\mu \partial_t\tilde{x}_\nu
-\partial_\sigma \tilde{x}_\mu \partial_\sigma \tilde{x}_\nu )
+4\pi i\alpha^\prime \tilde{b}^{\mu\nu }\partial_t\tilde{x}_\mu
\partial_\sigma \tilde{x}_\nu ],
\label{tildeS}
\end{equation}
where
\begin{eqnarray}
\tilde{g}^{\mu\nu }
&=&
((g-(2\pi\alpha^\prime )^2b^\prime g^{-1}b^\prime )^{-1})^{\mu\nu },
\nonumber
\\
\tilde{b}^{\mu\nu }
&=&
-(\frac{1}{g+2\pi\alpha^\prime b^\prime}b^\prime
\frac{1}{g-2\pi\alpha^\prime b^\prime})^{\mu\nu},
\nonumber
\\
\partial\tilde{x}_\mu
&=&
-g_{\mu\nu }\partial x^\nu -2\pi\alpha^\prime b^\prime_{\mu\nu }\partial
x^\nu ,
\nonumber
\\
\bar{\partial}\tilde{x}_\mu
&=&
g_{\mu\nu }\bar{\partial} x^\nu -2\pi\alpha^\prime
b^\prime_{\mu\nu }\bar{\partial}x^\nu .
\end{eqnarray}
The D$(p-\tilde{d})$-branes are mapped to D$p$-branes by this transformation
and
the vertex operator corresponding to the coordinate $X^\mu $ is mapped to
the vertex operator $\dot{\tilde{x}}_\mu $ for the gauge field on the
D$p$-branes
worldvolume. Therefore the effective action of the D$(p-\tilde{d})$-branes
can be
calculated as the dimensional reduction of the effective action for the
gauge field on the D$p$-branes in the background of $\tilde{g}^{\mu\nu}$ and
$\tilde{b}^{\mu\nu}$.

Thus we obtain the effective action for $X^\mu$:
\begin{equation}
S=\frac{1}{\tilde{g}_s(2\pi )^p(\alpha^\prime )^{\frac{p+1}{2}}}
Tr[\sqrt{\mbox{det}(\tilde{g}^{\mu\nu}
+2\pi\alpha^\prime \tilde{b}^{\mu\nu}
+i[X^\mu ,X^\nu ] )}] .
\nonumber
\end{equation}
Here $\tilde{g}_s$ should be determined so that the physical quantities
such as the brane tension should be invariant under T-duality.
As we saw in the previous sections, we can rewrite this action
as an action for noncommutative Yang-Mills theory by
substituting
\begin{equation}
X^\mu =C^{\mu\nu}(\hat{p}_\nu +\hat{a}_\nu ),
\end{equation}
We obtain the following Lagrangian after applying our rule eq.(\ref{momrule}):
\begin{equation}
\hat{\cal L}_\Phi
=
\frac{1}{G_s(2\pi )^p(\alpha^\prime )^{\frac{p+1}{2}}}
\sqrt{\mbox{det}(G+2\pi\alpha^\prime \hat{F}+\Phi )},
\label{gs}
\end{equation}
with
\begin{eqnarray}
\Phi
&=&
-C^{-1}-(2\pi\alpha^\prime )^2C^{-1}\tilde{b}C^{-1},
\nonumber
\\
G
&=&
-(2\pi\alpha^\prime )^2C^{-1}\tilde{g}C^{-1}.
\label{relations}
\end{eqnarray}
It is easy to show that the relations in eqs.(\ref{relations}) in addition
to $b=b^\prime +B$ implies eq.(\ref{phi}).

$G_s$ in eq.(\ref{gs}) should be determined so that the brane tension coincides
with the one for branes in constant background $g,b$ and we obtain:
\begin{equation}
G_s=g_s(
\frac{\mbox{det}(G+2\pi\alpha^\prime \Phi )}
{\mbox{det}(g+2\pi\alpha^\prime b)})^{\frac{1}{2}}.
\label{gseqn}
\end{equation}
For large $b$ with fixed $g,\Phi$, we get
$G_s\sim g_s(\mbox{det}b)^{\frac{1}{2}}, B\sim b$ which implies that
the noncommutative Yang-Mills coupling is
$g^2_{NC}\sim g_sb^2$ for D3-branes.

Thus we obtain gauge equivalent descriptions of the theory with
varying $\star$-products and they give the same effective actions
as the ones proposed by Seiberg and Witten. This fact strongly suggests that
our descriptions are physically equivalent to theirs.

\section{Conclusions and discussions}
\setcounter{equation}{0}

In this paper we have studied the correlation functions of
gauge invariant operators (Wilson loops) in noncommutative Yang-Mills.
Our investigation is based upon the equivalence of
noncommutative Yang-Mills and twisted reduced models.
Such a system appears in IIB matrix model as
infinitely extended D-brane solutions which preserve a part of SUSY.
We have studied correlation functions on the D-branes
in IIB matrix model through twisted reduced models.

Our conclusion is that there is a crossover in noncommutative
Yang-Mills theory at noncommutativity scale.
At long distances it reduces to ordinary Yang-Mills theory.
At large momentum scale, it becomes identical to large $N$ gauge theory.
The coupling of noncommutative Yang-Mills is shown to be
identical to the 't Hooft coupling in the high energy large $N$ gauge theory.
It immediately implies that the renormalizability of noncommutative
Yang-Mills is identical to large $N$ gauge theory.

We have pointed out that these high energy degrees of freedom
are not commutative to each other and correspond to nonlocal
degrees of freedom in IIB matrix model context.
We have explicitly constructed corresponding
Wilson loop operators.
For high energy modes, the noncommutativity of the backgrounds
does not matter and their dynamics is identical with
commutative backgrounds as long as the eigenvalues are uniformly distributed
in $\tilde{d}$ dimensions.
In fact each loop contribution from high energy modes can be estimated just
like the
quenched reduced models as
\beq
{g^2B^4}\sum_i f=g^2(2\pi )^2B^2
R^{\tilde{d}-4}\int^{\Lambda}_{\lambda} {d^{\tilde{d}}k\over (2\pi
)^{\tilde{d}}} f .
\eeq

As we have shown,
the crossover behavior
is consistent with the supergravity
description which is expected to be valid in the strong coupling regime.
The radius of $AdS_5$ is related to the coupling of noncommutative
Yang-Mills and it is also shown to be identical to 't Hooft coupling
of high energy large $N$ gauge theory.
We have also obtained gauge equivalent descriptions of the theory with
varying $\star$-products by introducing constant background
$g_{\mu\nu}$ and $b_{\mu\nu}'$.
We have shown that they give the same effective actions
as the ones proposed by Seiberg and Witten. Therefore their results can be
simply understood in our formalism.

In the context of IIB matrix model, strong coupling regime corresponds to
the high
density regime. Namely the spacing of the eigenvalues of the matrices
$R^4$ is small compared to $g^2$.
In this regime, it may be legitimate to expand the IIB matrix model
around D-branes since the classical action is small.
It appears that high density region represents weak gravity
and low density region represents strong gravity in IIB matrix model.
We would like to explore the intermediate density region.
For this purpose it is very desirable to formulate systematic high and
low density expansions of IIB matrix model.

\begin{center} \begin{large}
Acknowledgments
\end{large} \end{center}
This work is supported in part by the Grant-in-Aid for Scientific
Research from the Ministry of Education, Science and Culture of Japan.

\end{document}